\documentclass{emulateapj}
\newcommand{\uv}{\mbox{$u$-$v$}}

\newcommand{\Jb}{\mbox{Jy~beam$^{-1}$}}

\newcommand{\Ra}[4]{\mbox{${#1}^{\rm h} \; {#2}^{\rm m} \; {#3}\fs{#4} $}}
\newcommand{\dec}[4]{\mbox{${#1}\arcdeg \; {#2}\arcmin \; {#3}\farcs{#4} $}}

\shortauthors{Bartel and Bietenholz}
\shorttitle{SN1979C VLBI}
\begin{document}

\title{SHELL REVEALED IN SN~1979C}

\author{N. Bartel\altaffilmark{1} and M. F. Bietenholz\altaffilmark{1,2}}
\slugcomment{Received 2008 January 31; accepted 2008 April 14}
\journalinfo{{\em The Astrophysical Journal, 682, 2008 August 1}}
\altaffiltext{1}{Department of Physics and Astronomy, York University, Toronto,
M3J~1P3, Ontario, Canada}
\altaffiltext{2}{Hartebeesthoek Radio Observatory, PO Box 443, Krugersdorp, 
1740, South Africa} 


\begin{abstract}
VLBI observations at 5 GHz have revealed that supernova 1979C, in the
galaxy M100 in the Virgo cluster, has shell structure. The shell is
approximately circular with the 50\% contour deviating from a circle
by an rms of no more than 8\% of the radius.  The brightness
distribution along the ridge may vary. The position of the center of the shell is at
R.A.=\Ra{12}{22}{58}{66758} and decl.=\dec{15}{47}{51}{9695} (J2000), with a standard
error of 0.8 mas in each of the coordinates.  No isolated central
component is visible above a flux density limit of $\sim$150 $\mu$Jy which
corresponds to an upper limit of $\sim$15 times the corresponding spectral
luminosity of the Crab Nebula. The radio lightcurve is clearly falling again 
and the radio spectrum is now flatter than
at earlier times.  SN~1979C is only the fourth, and oldest (optically
identified) supernova of which a detailed image could be obtained.

\end{abstract}


\section{INTRODUCTION}

\objectname{SN~1979C} (type II-L) was discovered in the galaxy
\objectname{M100} \citep[\objectname{NGC~4321}, distance: 15.2 Mpc as
estimated by][]{Freedman+2001} in the Virgo cluster on 1979 April 19
by G. E. Johnson \citep{Mattei+1979} and subsequently became one of
the most optically and radio luminous Type II supernovae ever
found. On the basis of its radio lightcurve, its explosion date is
estimated to be 1979 April 4 \citep[see][]{Weiler+1986}, hereafter
$t=0$~yr. VLBI observations started as early as 1982. They resulted in
a determination of the angular expansion rate \citep{Bartel1985,
Bartel+1985a} and deceleration. This determination allowed an estimate
of the mass-loss to wind-velocity ratio of the progenitor, which was
found to be an order of magnitude lower \citep{SN79C} than that
reported from radio lightcurve fitting.

However, despite a quarter century of VLBI observations, the
relatively far distance to Virgo has prevented determining the
detailed structure of \objectname{SN~1979C} with VLBI\@. The
determination of the expansion rate and the deceleration had to be
based on size estimates made by using a model fit of the supernova
brightness distribution.  Clearly, a determination of the supernova's
morphology is important for a more accurate determination of the
expansion velocity and deceleration. Such more accurate determination
would also allow for a direct, geometric, estimate of the distance to
\objectname{SN~1979C} and Virgo using the expanding shock front method
\citep[ESM; see, e.g.,][]{SN93J-4} and resolve the ambiguity in an
earlier estimate \citep{SN79C}. It is also important for revealing the
morphological type of the supernova: purely shell-like or center
filled, perhaps with a bright compact component. Previously, only
three other optically-identified supernovae could be imaged with an
angular resolution high enough to unambiguously determine the
morphological type. All three, \objectname{SN 1986J}
\citep{Bartel+1991, SN86J-Sci}, \objectname{SN 1987A}
\citep{Gaensler+1997, Gaensler+2007}, and \objectname{SN 1993J}
\citep{Marcaide+1995a, BartelBR1995, SN93J-Sci, SN93J-3} have shell
structure, with SN~1986J having a central component in addition to the
shell. SN~1979C is the oldest optically-identified supernova that can
still be monitored with VLBI\@.  As such, it provides a bridge between
our understanding of supernovae (interacting with the circumstellar
medium left over from the progenitor) and supernova remnants
(interacting with the interstellar medium).  

Only a few supernova remnants which are less than a century old are
known, all extragalactic, e.g., in M82 \citep{McDonald+2001} and in
Arp220 \citep{Parra+2007}.  The youngest Galactic remnants are several
centuries old, e.g., the Crab Nebula, which shows a pulsar wind
nebula, Cas A, which shows a shell, and G21.5$-$0.9, which shows both
a pulsar wind nebula \citep[e.g.,][]{G21.5} and likely also a shell
\citep{MathesonS2005, Bocchino+2005}.  Since all radio supernovae
originate from the explosions of massive stars, they are expected to
leave a neutron star or a black hole.  Either a young neutron star or
a black hole may be accompanied by a bright, compact radio source near
the center of the expanding shell.  Young neutron stars are likely to
be associated with radio-bright pulsar wind nebulae, and black holes
may have radio bright jets due to an accretion-disk system.

Indeed, pulsar wind nebulae are found around a number of young
pulsars, although the youngest known pulsars have ages near 1000~yr.
A central radio source possibly associated with a neutron star or black hole was
recently found in SN~1986J \citep{SN86J-Sci}. Such a central radio
source might also be seen near the centers of other older supernovae
when clearly resolved images are obtained.  Here we report on such an
image of SN~1979C, discuss its morphology, radio lightcurve, and
spectral evolution, and search for signs of a central compact remnant
of the explosion.  The expansion rate, deceleration, and distance
determination will be reported elsewhere.

\section{OBSERVATIONS AND DATA ANALYSIS}

The VLBI observations were carried out on 2005 February 25 with a
global array of 21 antennas.  The array consisted of the Very Long
Baseline Array (VLBA; each of 25~m diameter), phased Very Large Array
(VLA; 130~m equivalent diameter), and the Robert C. Byrd telescope
(GBT; $\sim$105~m diameter), all of the National Radio Astronomy
Observatory, NRAO, in addition to the Arecibo Radio Telescope (305~m
diameter) and eight telescopes of the European VLBI Network, namely
the Effelsberg Radio Telescope, Germany
(100~m diameter), the Westerbork Array, The Netherlands (92~m
equivalent diameter), and the telescopes at Medicina and Noto, Italy
(32~m diameter), at Torun, Poland (32~m diameter), Jodrell Bank, 
Great Britain (26~m diameter), and Hartebeesthoek, South Africa (26~m
diameter).  We observed for a total of $\sim$15~h at a frequency of
5~GHz with a bandwidth of 32~MHz and recorded both senses of circular
polarization with a total recording bit rate of
256~Mbits~s$^{-1}$. The data were correlated with the NRAO VLBA
processor in Socorro and further analyzed with NRAO's Astronomical
Image Processing System (AIPS)\@.  The initial flux density
calibration was done through measurements of the system temperature at
each telescope, and then improved through self-calibration of the
reference sources.

We phase-referenced our VLBI observations to \objectname{J1215+1654},
which is 2.2\arcdeg\ away on the sky, and used, through most of the
run, a repeating cycle of 250 s, pointing for 170~s to SN~1979C and
for 80~s to the reference source.  Finally, we also used the VLA,
which was in the B configuration, to determine the total flux density
of SN~1979C at 1.43, 4.99, and 8.43~GHz.  In order to obtain the best
possible image, we phase self-calibrated the data after
phase-referencing.

\section{RESULTS}

\subsection{The Brightness Distribution}

In Figure~\ref{snf} we show the VLBI image of SN~1979C juxtaposed with
an earlier VLA image of the host galaxy M100. The supernova is located
in a southern spiral arm of the galaxy.  The VLBI image shows that the
supernova has shell structure
with a brightness minimum located only slightly to the west of the
center. The outer
contour at half of the maximum brightness is approximately circular
with rms deviations from circularity of 8\% of the radius.  We
think that this estimate is somewhat conservative since it is based on
an image convolved with an elliptical beam, which tends to elongate the
structure along the long axis of the beam, which is oriented
north-south. The 50\% contour is indeed slightly elongated in the
north-south direction. An ellipse fit to the 50\% contour has a ratio
between the major and minor axes of 1.10 and a p.a.\ of 173\arcdeg.

The maximum brightness is only 187 $\mu$\Jb\ and the rms noise
brightness in the background is 11~$\mu$\Jb.  How much could possible
deconvolution errors contribute to the uncertainty in the brightness
distribution of the supernova? We investigated the significance of the
appearance of the shell structure and of features in the brightness
distribution by using simulations.  We used a disk model as a test
source, converted it to visibility data for our \uv~coverage, added
noise, and then imaged and deconvolved the simulated data in the same
way as the observed data.  In this way we produced 36 images of our
model source with different noise realizations but with the rms
background noise brightness being always close to our observed background rms of
11~$\mu$\Jb. The average of the 36 images resembled the original disk
model, indicating that there are no systematic distortions or
biases. The individual images, however, varied in their appearance
with the rms of the brightness variations on the image being twice as large
as the rms of the background noise brightness. None of these images
had a closed ridge with a clear minimum in or near the center.  We
therefore exclude the possibility that the true brightness
distribution is uniform across the source or has a filled
center. Instead, with high probability, it is indeed a shell.

Our simulations also showed that the northern and southern
low-brightness extensions are likely due to corrugations in the image
plane caused by our incomplete \uv~coverage, and are therefore
probably not significant. The brightness varies along the ridge
between 187~$\mu$\Jb\ and 112~$\mu$\Jb, representing a
range of only $\pm 1.7\times$ the estimated on-source brightness
uncertainty of $\sim22$~$\mu$\Jb. It is not clear whether the variation is significant.

The brightness at the local minimum slightly to the west of
the center of the shell is 80~$\mu$\Jb.  The brightness increases
toward the east and is $\sim130~\mu$\Jb at the center, although no isolated
central component is visible. We determine an upper limit of the flux
density of any compact component in the central area of the shell of
$\sim150~\mu$\Jb.

\subsection{The Position}

\begin{figure}[ht]
\centering
\includegraphics[width=\textwidth, trim=0 0.in 0 0.in]{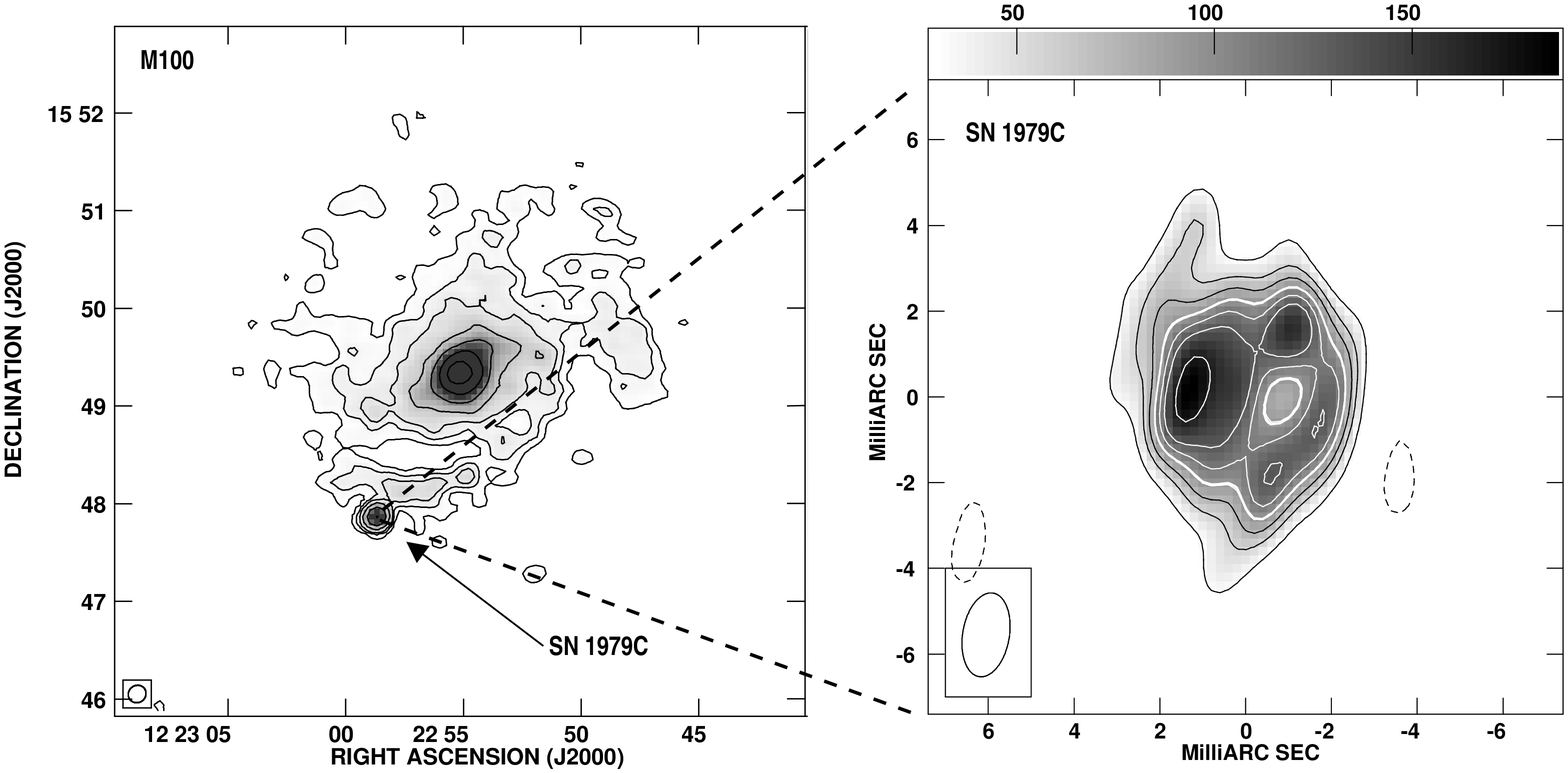}
\caption{Left side: A radio image of the spiral galaxy M100 (NGC~4321)
in the Virgo cluster, with SN~1979C in it, made with the
VLA in the C array at 1.7~GHz on 1996 March 15. The contour levels are at 1.3,
2.5, 4, 8, 16, 32, and 64\% of the peak brightness of 15.8~m\Jb. The full-width at half-maximum
(FWHM) restoring beam is plotted in the lower left corner. The image
is taken from \citet{SN79C}. Right side: the VLBI
image of SN~1979C, made at 5.0~GHz on 2005 February 25.  The contour
levels are at $-17$, 17, 30, 40, {\bf 50} (thick white contours), 60, 70, and 90\% of the peak
brightness of 187~$\mu$\Jb. The grey scale, in $\mu$\Jb,
is given at the top.  The rms brightness of the background noise is 11
$\mu$\Jb. The FWHM of the restoring beam is 2.0~mas $\times$ 1.1~mas
oriented at a p.a.\ of $-10$\arcdeg. North is up and east to the
left. The origin of the coordinate system is at R.A.=\Ra{12}{22}{58}{66759} 
and decl.=\dec{15}{47}{51}{9695} (J2000) with a standard error of 0.8 mas in each coordinate. It is
at the approximate center of the outer 50\% contour (see text).}
\label{snf}
\end{figure}

Phase-referencing allowed us to determine the position of SN~1979C
relative to that of our calibrator, J1215+1654. The calibrator's
position was taken from the VLBA Calibrator Survey
\citep[VCS1,][]{Beasley+2002} as R.A.=\Ra{12}{15}{03}{979123} and
decl.=\dec{16}{54} {37}{95733} (J2000). The standard
error\footnote{\citet{Beasley+2002} call this error ``inflated formal
error.''} is 0.5 and 0.6~mas in RA and decl., respectively. In our
VLBI image of SN~1979C in Figure \ref{snf}, we place the origin of
the coordinate system at the estimated center of the outer 50\%
contour.  This coordinate origin is at the position of
R.A.=\Ra{12}{22}{58}{66758} and decl.=\dec{15}{47}{51}{9695} (J2000).  We
estimate that our standard error of the position of SN~1979C relative
to J1215+1654 is 0.5~mas in each coordinate \citep[compare
with][]{PradelCL2006}. 
Adding in quadrature the standard position error of J1215+1654's to
that of SN~1979C, we estimate a standard error of the position of the
center of the image in Figure~\ref{snf} to be 0.8~mas in each
coordinate.

As a check, we also determined the position of SN~1979C relative to
J1215+1654 by using our earlier 1.7~GHz data from 2001. At this
frequency, SN~1979C was barely resolved, and contributions to the
position error due to unmodeled effects of the ionosphere expected to
be much larger. We estimate a standard error of 2~mas in each
coordinate.  Our estimated relative position of SN~1979C at 1.7~GHz in
2001 was within 1~mas of the more accurate position given above, well
within the uncertainties.

\subsection{VLA Total Flux Densities}
We determined the total flux density, $S_{\nu}$, of SN~1979C from the
interferometric observations with the VLA made during the time of our
VLBI observations. We list these flux densities for frequencies,
$\nu$, of 8.43, 4.99, and 1.43~GHz in Table~\ref{fluxt}. In addition, we
give the flux densities for $\nu = 8.46$ and 1.67~GHz measured in 2001, for
comparison. Then we convert the flux densities to $\nu = 4.89$ and 1.47~GHz and plot
them in Figure~2, together with the much larger set of
flux density determinations at these frequencies of \citet{Weiler+1986, Weiler+1991},
\citet{Montes+2000}, and \citet{SN79C}.

\begin{deluxetable*}{l c c c c c@{\hspace{0.5in}}c}  
%
%
\tabletypesize{\small}
\tablecaption{Total flux densities and spectral indices of SN~1979C}
\tablehead{ 
\colhead{Date} & \colhead{Age (yr)\tablenotemark{a}}  & \multicolumn{4}{c}{Flux 
densities (mJy)\tablenotemark{b}} 
& \colhead{$\alpha$\tablenotemark{c}\phn\phn}\\
& & \colhead{1.43 GHz} &
    \colhead{1.67 GHz}   & \colhead{4.99 GHz}  & {8.43 GHz}
}
\startdata
 2001 Feb 23& 21.89 &                 & 4.28$\pm$0.23 &                 & \phn2.3$\pm$0.4\tablenotemark{d}\tablenotemark{e} & $-0.38\pm0.15$ \\
 2005 Feb 25& 25.90 & 3.19$\pm$0.22   &               & 1.68$\pm$0.09   & 1.37$\pm$0.10                                 & $-0.49\pm0.09$ \\     
\enddata
\tablenotetext{a}{Time since assumed explosion date of 1979 April 4 (1979.26).}
\tablenotetext{b}{Total flux density, $S_\nu$, measured with the VLA\@.
The uncertainties are approximate standard errors and were computed by adding 
in quadrature the statistical standard error and a systematic calibration 
error of $\ge 5$\%.} 
\tablenotetext{c}{The spectral index, $\alpha$, determined from  
the flux density measurements, $S_{\nu}$, at frequency, $\nu$:
 $S_{\nu}\propto\nu^{\alpha}$. For our last epoch a weighted fit was used. The errors 
are one standard deviation, determined by varying the flux density values by their standard errors.}
\tablenotetext{d}{Data are from \citet{SN79C}.}
\tablenotetext{e}{The observing frequency was 8.46 GHz.}

\label{fluxt}
\end{deluxetable*}

The new measurements of the flux density show that the supernova
is again fading, reaching at 1.5~GHz its lowest level ever.  The bumpy
pattern seen previously is extended downward.

\subsection{Radio Spectra}

\citet{Montes+2000} found that the optically-thin spectrum had a
spectral index, $\alpha$, of $-0.75$ with a deviation range of $-0.76$
to $-0.63$ for $t<12$~yr (i.e., before 1991; $\alpha$ being obtained
from the fit of an empirical model to the multi-frequency radio
lightcurves), and $\alpha \sim -0.70$ for $t\geq12$~yr.  At their
latest times, $t=17.7$ and 18.8~yr, they obtained $\alpha =
-0.63\pm0.03$.  Our latest measurements give still flatter spectra
(see Table~\ref{fluxt}).  The difference of our spectral indices to
that of $-0.63$ may be significant. In Figure~\ref{spectraf} we plot
our flux density measurements and the corresponding spectra for our
two latest epochs. For comparison we show the flux density
measurements at $t=17.7$ and 18.8~yr from \citet{Montes+2000} and the
corresponding average spectrum.  Clearly, there is evidence that the
spectrum of SN~1979C has flattened further since $t\sim$17~yr.

\section{DISCUSSION}

SN~1979C is the fourth (optically identified) extragalactic supernova
for which an image with sufficient angular resolution could be made to
determine the nature of the brightness distribution. The other three, 
as mentioned in the introduction, are SN 1986J, SN 1987A,
and SN 1993J\@.  Our VLBI image shows that SN~1979C has a morphology
similar to the other three, namely a fairly to highly round shell,
with the brightness possibly being modulated along the ridge. The
possible brightness modulation likely arises from density
irregularities either in the circumstellar medium or in the ejecta.

The flux density of SN~1979C is now clearly again on the downward
trend.  Earlier, it was suggested that the radio lightcurve exhibits a
quasi-periodic pattern \citep{Weiler+1992} which was interpreted as
possibly being due to the dynamical modulation of the wind from the
red supergiant progenitor by a binary companion
\citep{SchwarzP1996}. With more data up to $t\sim19$~yr, it was
noticed that any periodic variation of the radio lightcurve was not
extended beyond $t\sim10$~yr.  Instead, it was suggested that, with
the flattening of the radio lightcurve, the supernova had apparently
entered a new stage of evolution \citep{Montes+2000}.

With our new data up to $t\sim26$~yr, which show that the lightcurve
is again falling after $t \sim 16$~yr, a new look at the
characteristics of the radio lightcurve is warranted. In particular,
we think that the evidence for a periodicity in the radio lightcurve
or a new stage of evolution now seems weaker.  If at all, then the
radio lightcurve appears to be periodic on a logarithmic time scale.
However, a closer inspection leads us to think that, rather than being
periodic on a logarithmic time scale, the radio lightcurve is
modulated on a whole range of time scales. For instance, there appear
to be variations on time scales of less than a year. There are also
variations on time scales of several years with e.g., a 5-yr bump in
the radio lightcurve at $t\sim7$~yr and a bump lasting at least 15 yr
at the end of the observed radio lightcurve.  These variations are
likely caused by variations in the density of the ejecta \citep[for
discussions in the case of SN~1993J, see][]{MioduszewskiDB2001} and/or
of the circumstellar medium. In case of the variations in the latter,
the radio lightcurve can be thought of as a ``time machine''
which records in reverse the mass-loss to wind-velocity ratio of the
progenitor for thousands of years before the star died.

The spectrum appears to have flattened somewhat with the spectral
index being above the deviation range of $-0.76$ to $-0.63$ given by
\citet{Montes+2000}. A similar behavior is also seen for SN 1993J
where the spectrum flattens while the source dimming
accelerates\footnote{Please note that although \cite{Weiler+2007}
claim that the spectral index of SN~1993J remains constant over time,
an increase of the spectral index consistent with the measurements of
\citet{SN93J-2} and \citet{Perez-TorresAM2002} can be clearly seen in
their data with the smallest errors and/or scatter.} \citep{SN93J-2,
Perez-TorresAM2002}.

No clearly isolated component was found in the central region of the
projected shell of SN~1979C\@. However, the brightest portion of the
shell appears to extend toward the center in projection,
and could overlap with any compact component that may be present in
the interior.  An isolated central component was in fact found for
SN~1986J, with a spectrum peaking at $\sim$20~GHz while the emission
from the remaining parts of the supernova has an optically thin
spectrum from at least 1.7 to 22~GHz \citep{SN86J-CosparII,
SN86J-Sci}. This isolated central component may be emission
associated with the compact remnant of the explosion.

It is intriguing that the spectrum of SN~1979C shows evidence of
flattening with time. Such flattening would be expected for a
supernova with emission from a compact remnant that is starting to
appear in the center of a shell which is expanding and becoming
increasingly transparent at progressively higher frequencies.
However, no central component can clearly be seen in SN~1979C, so the
observed spectral flattening cannot be ascribed to such a component.

It is probable, therefore, that the spectral flattening is intrinsic
to the shell's synchrotron emission.  The flattening must then be due
to a change in the population of accelerated particles which produce
the synchrotron emission.  Calculations have shown that physical
conditions in supernova shocks can effect the spectral index of radio
emission \citep[e.g.,][]{Ellison+2007, EllisonBB2000}.

Our new upper limit on the flux density of any central compact source
in SN~1979C of 150~$\mu$Jy puts strong constraints on the luminosity
of such a source. \citet{BandieraPS1984} argued that young
pulsar wind nebulae would have spectral luminosities of 10 to 1000
times that of the Crab Nebula.
The spectral luminosity of the central component of SN 1986J is
$\sim$200 times that of the Crab Nebula between 14~GHz and 43~GHz
\citep{SN86J-CosparII}.  In contrast, the upper limit on the 5~GHz
spectral luminosity of any central component in SN~1979C is only 15
times the corresponding spectral luminosity of the Crab Nebula, more
than an order of magnitude fainter than both that in SN~1986J and our
previous limit for SN~1979C \citep{NB-COSPAR}, and at the low end of
\citet{BandieraPS1984}'s estimate. Any pulsar wind nebula that may
exist inside the shell can probably not yet be detected because the
shell is likely still opaque at radio frequencies.  Since the
supernova shell becomes less dense and more transparent as it expands,
any central component will become more easily detectable in the
future. Searching for emission from the compact central remnant of the
explosion therefore remains an intriguing prospect.

\acknowledgements

Research at York University was partly supported by NSERC\@.  NRAO is
a facility of the National Science Foundation (NSF) operated under
cooperative agreement by Associated Universities, Inc. The Arecibo Observatory is
part of the National Astronomy and Ionosphere Center, which is
operated by Cornell University under a cooperative agreement with the
NSF\@. The European VLBI Network is a joint facility of European and
Chinese radio astronomy institutes funded by their national research
councils.  We have made use of NASA's Astrophysics Data System
Abstract Service.

\bibliographystyle{apj}
\bibliography{/home/bartel/paper/nb-bib}


\clearpage

\begin{figure}[ht]
\includegraphics[height=3.5in]{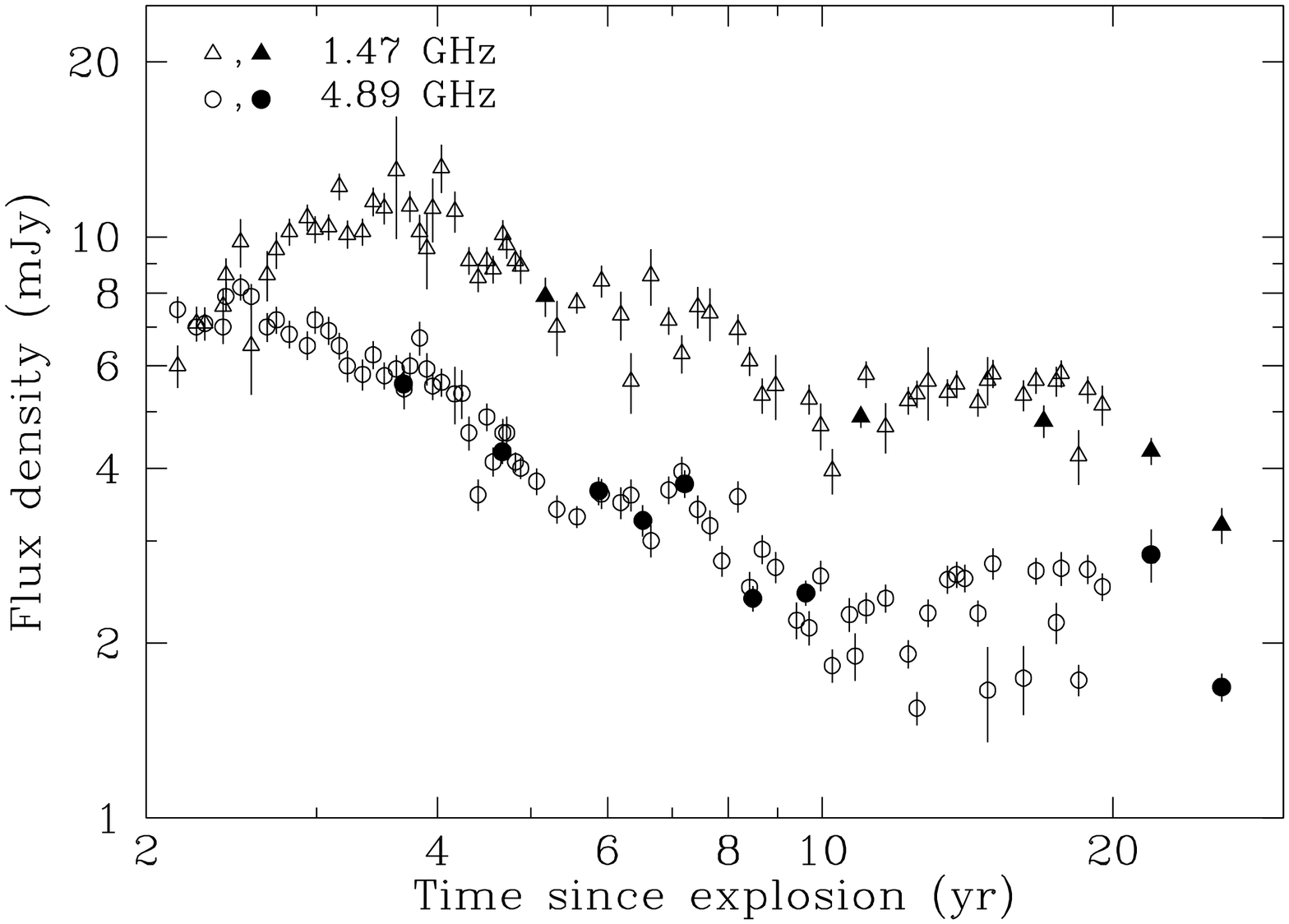} \figcaption{The radio lightcurves
at 1.47 and 4.89~GHz.  Open symbols represent values taken from
\citet{Weiler+1986, Weiler+1991} and \citet{Montes+2000}. Filled symbols
represent values from \citet{SN79C} and this paper.
The observed values were taken at a number of slightly different
frequencies. If necessary, they were converted to the above frequencies with
$\alpha=-0.75$ \citep{Montes+2000}, except for our values at $t=22$
and 26~yr, for which we used the spectral indices from
Table~\ref{fluxt}. The 4.89-GHz value at $t=22$~yr is interpolated
between the 1.67 and 8.43~GHz values from Table~\ref{fluxt}.}

\bigskip
\bigskip
$^3$~\footnotesize{
Conversion with a slightly different spectral index of $-0.63$, as
used in \citet{SN79C} would result in values virtually
indistinguishable from those plotted.}
\label{fluxf}
\end{figure}

\begin{figure}[ht]
\centering
\includegraphics[height=3.5in]{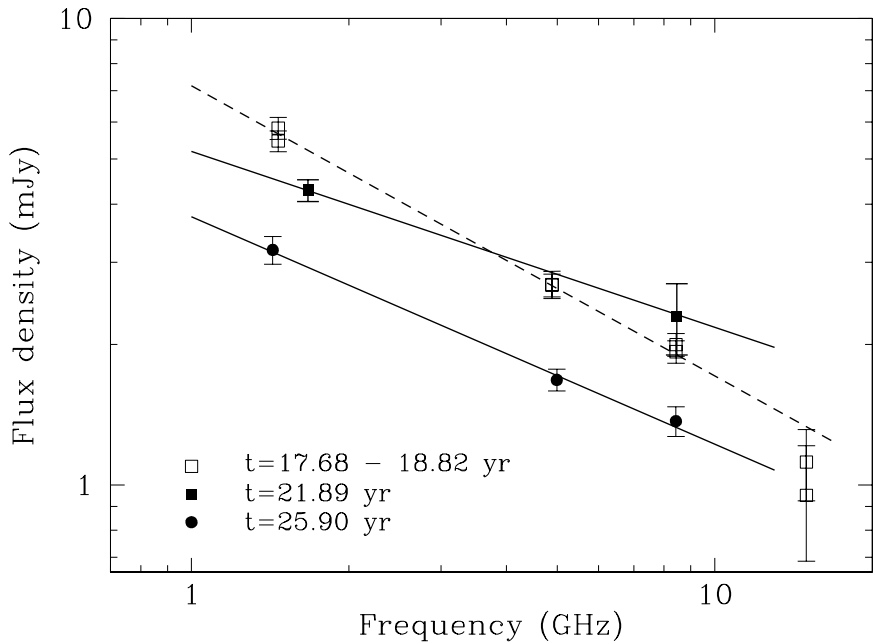} 
\caption{The flux densities from Table~\ref{fluxt} for the epochs
given in the lower left.  For comparison and to show evidence for the flattening of
the spectrum with time, the flux densities given by
\citet{Montes+2000} for the time range from $t=17.68$ to 18.82 yr after
the explosion are also plotted. The straight lines indicate fits
to the corresponding data.}
\label{spectraf}
\end{figure}

\end{document}